\documentclass[12pt,preprint]{aastex}

\shorttitle{CH$_3$D in L1527}
\shortauthors{Sakai et al.}

\usepackage{lscape}
\begin{document}


\title{TENTATIVE DETECTION OF DEUTERATED METHANE TOWARD THE LOW-MASS PROTOSTAR IRAS 04368+2557 IN L1527}


\author{Nami Sakai\altaffilmark{1, 2}, Yancy L. Shirley\altaffilmark{3}, Takeshi Sakai\altaffilmark{4}, Tomoya Hirota\altaffilmark{5}, Yoshimasa Watanabe\altaffilmark{1}, and Satoshi Yamamoto\altaffilmark{1, 2}}
\altaffiltext{1}{Department of Physics, The University of Tokyo, Bunkyo-ku, Tokyo 113-0033, Japan}
\altaffiltext{2}{Research Center for the Early Universe, The University of Tokyo, Bunkyo-ku, Tokyo 113-0033, Japan}
\altaffiltext{3}{Steward Observatory, University of Arizona, 933 N. Cherry Ave. Tucson, AZ 85721, USA}
\altaffiltext{4}{Institute of Astronomy, The University of Tokyo, Osawa, Mitaka, Tokyo, 181-8588, Japan}
\altaffiltext{5}{National Astronomical Observatory of Japan, Osawa, Mitaka, Tokyo 181-8588, Japan}

\begin{abstract}
The millimeter-wave rotational transition line ($J_K = 1_0 - 0_0$) of deuterated methane CH$_3$D has tentatively been detected toward the low-mass Class 0 protostar IRAS 04368+2557 in L1527 with the Heinrich Hertz Submillimeter Telescope.   This is the first detection of CH$_3$D in interstellar clouds, if confirmed.  The column density and fractional abundance of CH$_3$D are determined to be $(9.1\pm3.4) \times 10^{15}$~cm$^{-2}$ and $(3.0\pm1.1) \times 10^{-7}$, respectively, where we assume the rotational temperature of 25 K. The column density and fractional abundance of the gaseous CH$_4$ are estimated to be $(1.3-4.6) \times 10^{17}$~cm$^{-2}$ and $(4.3-15.2) \times 10^{-6}$, respectively, by adopting the molecular D/H ratios of 2--7~\% reported for various molecules in L1527.  The fractional abundance of CH$_4$ is higher than or comparable to that found in high-mass star-forming cores by infrared observations.  It is sufficiently high for triggering the efficient production of various carbon-chain molecules in a lukewarm region near the protostar, which supports the picture of the warm carbon-chain chemistry.
\end{abstract}

\keywords{ISM: individual (L1527) ISM: molecules}


\section{Introduction}

Methane (CH$_4$) is the most fundamental hydrocarbon molecule in space, and is thought to play an important role in interstellar chemistry and planetary chemistry.  The gas phase CH$_4$ has been identified toward bright infrared sources like high-mass young stellar objects by observing the infrared vibration-rotation spectra of the $\nu_3$ (C-H stretching: 3~$\mu$m) and $\nu_4$ (CH$_2$ bending: 7~$\mu$m) bands \citep{lac91, boo04, kne09}.  Such observations are also carried out toward comets, planets, and satellites like Titan \citep[e.g.][]{pen05}.  Furthermore, CH$_4$ is known as one of the major constituents of icy grain mantles in interstellar clouds.  The infrared absorption of the $\nu_3$ ($3.3$~$\mu$m) and $\nu_4$ ($7.7$~$\mu$m) bands have been observed toward various sources including low-mass protostars \citep[e.g.][]{lac91, obe08, obe11}.   In spite of this progress, our understanding of the distribution and the abundance of CH$_4$ in the gas phase is still limited particularly for cold interstellar medium including low-mass star-forming regions.  This is because CH$_4$ has no permanent dipole moment, and does not emit the pure rotational spectral lines which can readily be observed with radio telescopes.  Furthermore, the low-mass protostars are not bright enough for high-resolution infrared absorption spectroscopy.  

One practical way to overcome this situation is to observe its mono-deuterated species, CH$_3$D, which emits the pure rotational spectral lines due to its small electric dipole moment (0.005657~D) \citep{wof70, wat79}.  However, the past efforts to detect the $J_K = 1_0 - 0_0$ line (232.6~GHz) of CH$_3$D were not successful even toward Orion-KL \citep{pic80, wom96}.  Since active high-mass star-forming regions contain many kinds of saturated complex organic molecules, their contamination make it difficult to detect the line.  At that time, it was thought that detection of the CH$_3$D line would be difficult in low-mass star-forming regions because of its relatively low column density in comparison with high-mass star-forming regions.

Recently, it was recognized that CH$_4$ plays a crucial role in chemical processes occurring in low-mass star forming regions.  Sakai et al. (2008; 2009a) found low-mass star-forming regions which harbor extremely rich carbon-chain molecules.  They are L1527 in Taurus and IRAS15398-3359 in Lupus.  According to the interferometric observation toward L1527, carbon-chain molecules and their related species show a steep increase in abundance inward of 500--1000~AU \citep{sak10}, where the gas kinetic temperature is higher than 20~K \citep{shi02, jor02}.  Furthermore, they reside even in the gas infalling to the protostar.  In order to explain the observational results, it is proposed that carbon-chain molecules are regenerated near the protostar triggered by evaporation of the solid CH$_4$ on dust grains (Warm Carbon Chain Chemistry: WCCC) \citep{sak08}.  The desorption temperature of CH$_4$ is about 25~K \citep{col04}, which is lower than that of H$_2$O ($\sim 100$~K).  Hence CH$_4$ can be evaporated in a lukewarm region ($20-30$~K) around a newly born protostar.  The CH$_4$ evaporated from dust grains reacts with C$^+$ to form C$_2$H$_3^+$, which gives C$_2$H$_2$ and C$_2$H by electron recombination reactions.  Further reactions of C$_2$H$_2$ and C$_2$H with C$^+$ produce longer carbon-chain molecules.   A basic part of this scheme is confirmed by chemical model calculations \citep{aik08, has08, har08}.  The WCCC picture is consistent with the observed distribution of the carbon-chain molecules in L1527 \citep{sak10}.

Because of a high expected abundance of CH$_4$, the WCCC sources would be a good target to search for the CH$_3$D line.  The molecular D/H ratios are measured to be consistent for various molecules toward L1527 \citep{sak09b}, and hence, we can make a reasonable estimation for the abundance of gaseous CH$_4$ from the observation of CH$_3$D.  The result will verify the WCCC mechanism, and will lead us to a deeper understanding of a role of CH$_4$ in interstellar chemistry.  With this motivation, we conducted a sensitive observation of the CH$_3$D line toward L1527.\\ 
 
\section{Observations}

We first observed the $J_K = 1_0 - 0_0$ line (232.644301~GHz) of CH$_3$D toward the low-mass protostar, IRAS 04368+2557 in L1527 with the IRAM 30 m telescope in April 2008.   The observed position was $(\alpha_{2000}, \delta_{2000}) = (04^{\rm h} 39^{\rm m} 53^{\rm s}.87, 26^{\circ} 03^{\prime} 09^{\prime\prime}.7)$.  We employed the A230/B230 receiver as a frontend.  The beam size is 10.$^{\prime\prime}$6.  In this observation, we found a hint of the CH$_3$D line at the right $V_{\rm LSR}$ velocity with the intensity of 20~mK ($T_{\rm MB}$), although the confidence level is only 2.3~$\sigma$.

Encouraged by this result, we then conducted a long integration observation of this line toward the same position with Heinrich Hertz Submillimeter Telescope (HHT) \footnote{The HHT is operated by the Arizona Radio Observatory (ARO), Steward Observatory, University of Arizona.} in January 2011 and February 2012.  The ALMA Band 6 ($1.3-1.1$)~mm prototype receiver was used as a frontend in the 4 IF dual polarization mode, whose system noise temperatures ranged from 180~K to 250~K.  In the observation in 2011, the CH$_3$D $J_K = 1_0 - 0_0$ line was placed in the lower sideband with the CS $J=5-4$ line (244.9355565~GHz) in the upper sideband.  The sideband rejection was about -13~dB.  The beam size and the main beam efficiency of the telescope are listed in Table \ref{tab:para}.  In the observation in 2012, the CH$_3$D $J_K = 1_0 - 0_0$ line was placed in the upper sideband with the C$^{18}$O $J=2-1$ line (219.5603541~GHz) in the lower sideband.  The telescope pointing was checked every 1.5 hours by observing the continuum emission of Jupiter, or the CS($J=5-4$) emission of S231 and IRC$+$10216, and was maintained to be better than 10$^{\prime\prime}$.  A small daily variation of the intensity was calibrated by using the CS or C$^{18}$O line.  The backend used was filterbanks whose bandwidth and resolution are 64~MHz and 250~kHz, respectively.  The frequency resolution corresponds to the velocity resolution of 0.32~km s$^{-1}$, which is comparable to or slightly smaller than the line width in this source (0.3-0.6~km s$^{-1}$: Sakai et al. 2008; 2009).   The observation was made in position-switching mode with the off-position of $\Delta \alpha =20$~arcminutes.  The final spectrum was placed on the $T_{\rm MB}$ scale using observations of Jupiter.\\

\section{Results}
\begin{figure}
\epsscale{.70}
\plotone{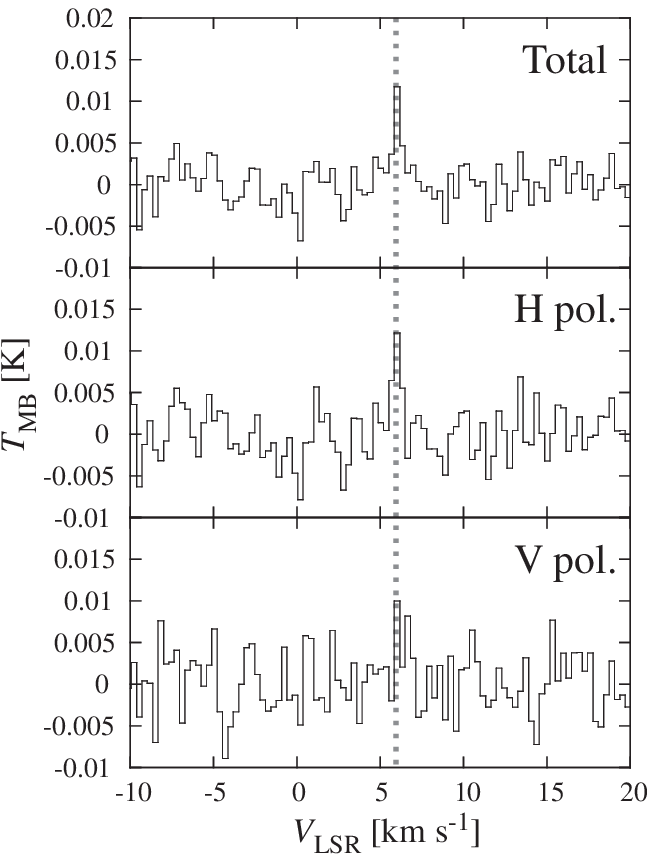}
\caption{Spectral line profile of  CH$_3$D ($J_K = 1_0 - 0_0$) observed toward L1527 with HHT.  The total spectrum is the weighted sum of the V- and H-polarization spectra.  The V-polarization spectrum shows a larger noise level than the H-polarization spectrum, because we have to exclude apparent bad-scan data caused by the receiver instability in integration of the V-polarization data.  As a usual reduction process, each 5~minute integration scan were carefully checked to eliminate the bad scan data with a heavy baseline distortion and/or an apparent strong noise spikes.  Note that no apparent spikes appear at the line channel.  The on-source integration time for the H and V spectra are 25 and 13 hours, respectively.  The vertical dotted line shows the averaged $V_{\rm LSR}$ value reported for deuterated species in this source (5.9~km~s$^{-1}$) \citep{sak09b}.  \label{fig:ch3d}}
\end{figure}

\begin{figure}
\epsscale{1.0}
\plotone{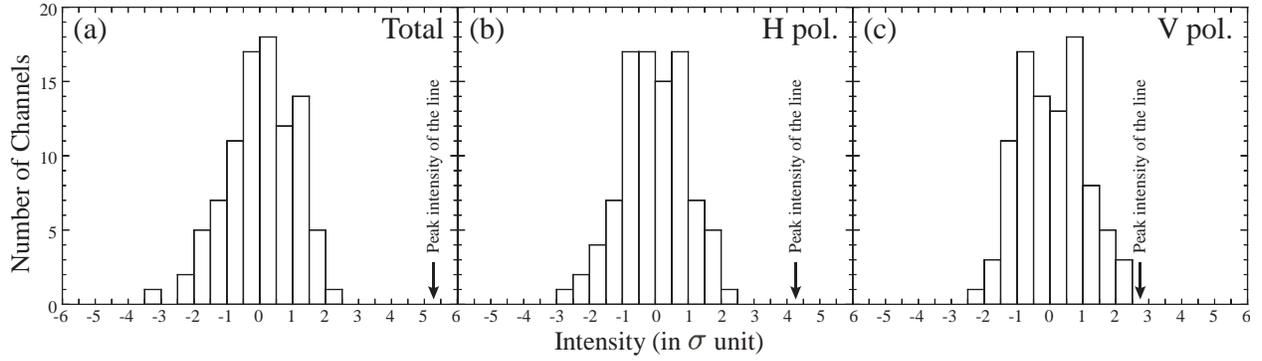}
\caption{Noise distributions of the spectra shown in Figure~\ref{fig:ch3d}.  Panels (a), (b), and (c) represent the noise distributions for the total, H-polarization, and V- polarization spectra, respectively.  This figure is prepared from the data for the central 93 channels excluding the line channels, where the baseline of the spectrum is subtracted.  The distribution is close to the Gaussian form.  The arrows indicate the peak intensities of the line.  It is significantly detected for the total and H-polarization spectra. \label{fig:gau}}
\end{figure}

In the 2011 observation with HHT, we tentatively detected the $J_K = 1_0 - 0_0$ line of CH$_3$D toward the low-mass protostar IRAS 04368+2557 in L1527, as shown in Figure~\ref{fig:ch3d}.  The line can be recognized in the both (H and V) polarization data, although it is marginally seen in the V polarization data due to a larger noise level.   The line is not well resolved due to a limited frequency resolution of the backend employed.  In preparation of the spectra, each 5~minute integration scan were carefully checked to eliminate the bad scan data with a heavy baseline distortion and/or an apparent strong noise spikes, as a usual data reduction process.  Note that no apparent spikes appear at the line channel.  For the total spectrum, the confidence level of the detection is 7.9~$\sigma$ in the integrated intensity, and 5.2~$\sigma$ in the peak intensity.  This line shows the strongest intensity in the observed band (128 spectral channels), and the noise reveals a Gaussian distribution, as shown in Figure~\ref{fig:gau}.  Hence, the line is significantly detected in the total spectrum.  The noise distributions of the H and V polarization spectra are also shown in Figures~\ref{fig:gau}b and \ref{fig:gau}c, respectively.  They show the Gaussian form, although the V polarization data have a slightly poor shape.

To confirm the line, we also carried out the observation with the different frequency setting in 2012, where the CH$_3$D line is set in the upper sideband, and marginally confirmed the line with the 3.6~$\sigma$ confidence level.  However, we unfortunately suffered from serious spurious signals probably due to instability of the receiver and the backend in the 2012 observation in contrast to the 2011 observation, and hence, we did not use the 2012 data for the final spectrum.
 
The linewidth and the LSR velocity are consistent with those of other carbon-chain molecules observed in this source (e.g. Sakai et al. 2008; 2009b).  We carefully checked the molecular line databases such as CDMS \citep{mul05}, JPL Catalog \citep{pic98}, and confirmed that no appropriate line exists at the frequency of $\pm$0.5~km s$^{-1}$ except for the CH$_3$D line.  We also confirmed no strong line in the image band.  The lines of saturated complex organic molecules like HCOOCH$_3$, C$_2$H$_5$CN, and (CH$_3$)$_2$O, which give congested spectra, are not detected in L1527 even with a very sensitive observation \citep{sak07}.  This is in contrast to the high mass star-forming regions like Orion KL, and also to the low-mass star-forming regions with the hot corino activity (e.g. IRAS16293$-$2422; Cazaux et al. 2003).  Furthermore, the linewidth in L1527 is narrower than those in Orion KL ($\sim$5~km s$^{-1}$) and IRAS16293$-$2422 ($\sim$2~km s$^{-1}$).  Although L1527 harbors long various carbon-chain molecules and their isomers, their rotational transitions in the 200~GHz region generally have very high upper state energies ($>$100 cm$^{-1}$, for carbon-chains with for or more carbon atoms) and are difficult to be excited in L1527.  Therefore, an accidental matching of other lines would be unlikely, and our identification of this line to CH$_3$D is the most reasonable.

The column density of CH$_3$D is estimated to be $(9.1\pm3.4) \times 10^{15}$~cm$^{-2}$ from the integrated intensity by assuming the LTE (local thermodynamic equilibrium) condition with the excitation temperature of 25~K, which corresponds to the evaporation temperature of CH$_4$ from dust grains.  The source size is assumed to be 20$^{\prime\prime}$ \citep{sak10}.  Here the error is calculated from three times the standard deviation of the integrated intensities.  The fractional abundance of CH$_3$D is evaluated to be $(3.0\pm1.1) \times 10^{-7}$, where the H$_2$ column density is assumed to be $3 \times 10^{22}$~cm$^{-2}$ \citep{jor02}.  Temperature dependences of the column density and the fractional abundance are carefully inspected, as summarized in Table \ref{tab:column}.  The column density and the fractional abundance vary roughly linearly with the temperature over the range from 15 to 35~K.  In this temperature range, the difference is within a factor of 2. 

varies within a factor of 2.  The CH$_3$D column density is much less than the upper limit reported for Orion-KL ($N < 6 \times 10^{18}$~cm$^{-2}$) by three orders of magnitude \citep{wom96}.  When we assume the deuterium fractionation ratio of 2--7~\% by referring the ratios for various carbon-chain molecules observed toward L1527 \citep{sak09b}, the CH$_4$ column density is evaluated to be (1.3--4.6) $\times 10^{17}$~cm$^{-2}$ by using the CH$_3$D column density for the excitation temperature of 25~K.  Hence, the fractional abundance of CH$_4$ is (4.3--15.2) $\times 10^{-6}$.  In this calculation, we do not explicitly consider the statistical factor due to four equivalent hydrogen atoms in CH$_4$, because the deuterium fractionation ratios observed in L1527 are within the above range regardless of the number of equivalent hydrogen atoms (i.e. 0.05, 0.03, 0.07, and 0.04 for l-C$_3$D/l-C$_3$H, C$_4$HD/C$_4$H$_2$, c-C$_3$HD/c-C$_3$H$_2$, and NH$_2$D/NH$_3$, respectively).\\

\section{Discussions}

In this study, we tentatively detected the rotational spectral line of CH$_3$D in interstellar space for the first time.  Although no cataloged lines  do not match to the observed line, it might be a line of unknown species.  To rule out this possibility completely, we need to observe the other rotational transitions.  The intensities of the $J_K = 2_0 - 1_0$ (465~GHz; $E_u=33$~K) and $3_0 - 2_0$ (698~GHz; $E_u=67$~K) lines are predicted to be 21~mK and 12~mK by assuming the excitation temperature of 25 K and the linewidth of 0.5~km s$^{-1}$.  However, their sensitive observations are difficult because of the relatively heavy atmospheric absorption at these frequencies.  We have therefore decided to publish the present result as the tentative detection of CH$_3$D, considering its importance in astrochemistry.

The gas-phase CH$_4$ abundance has been reported for a few high-mass star-forming regions by observing the infrared absorption lines.  Lacy et al. (1991) found that the gas-phase CH$_4$ abundance is typically 10$^{-3}$ times the CO abundance.  If we assume the fractional abundance of CO to be 10$^{-4}$, this corresponds to the fractional abundance of 10$^{-7}$.  This value is lower than that observed in the present study by an order of magnitude.  On the other hand, Boogert et al. (2004) reported the gas-phase CH$_4$ abundance to be higher than $3 \times 10^{-6}$ toward the core of NGC7538~IRS9, and Knez et al. (2009) recently reported it to be $4.8 \times 10^{-6}$ toward NGC7538~IRS1.  These abundances are comparable to our result, although their observation traces much hotter region ($T_{\rm rot}$=55--674~K) than our present observation.

According to the chemical model calculation of dynamically evolving cores by Aikawa et al, (2008), CH$_4$ is mainly produced in grain mantles, and it is evaporated in a lukewarm region near the protostar.  The gas-phase CH$_4$ abundance in L1527 comparable to or slightly lower than that expected from their chemical model calculation ($10^{-5}$).  It should be noted that the fractional abundance of CH$_3$D is also comparable to or slightly lower than the result of their new chemical model calculation involving the deuterated species ($\sim 3.6 \times 10^{-7}$) \citep{aik12}.  If the emitting region of the CH$_3$D line is smaller than 20$^{\prime\prime}$, the abundance of CH$_3$D becomes higher, giving a better agreement with the model result.  

In this relation, it is interesting to compare our result with the solid CH$_4$ abundance.  The infrared absorption spectrum of solid CH$_4$ is not available for L1527, because the protostar is heavily obscured even in the mid-infrared region.  However, it is detected in another WCCC source, IRAS15398-3359 \citep{obe08, sak09a}.  The abundance of the solid CH$_4$ relative to the H$_2$O ice is reported to be 0.06.  When we assume the fractional abundance of the H$_2$O ice relative to the gas-phase H$_2$ to be 10$^{-4}$, the fractional abundance of the solid CH$_4$ is estimated to be $6 \times 10^{-6}$.  This is almost comparable to the gas-phase CH$_4$ abundance found in L1527.  

Recent observational studies of low-mass star-forming regions reveal a significant chemical diversity among the low-mass Class~0 protostars.  Two distinct cases are hot corinos and the WCCC sources.  The hot corino chemistry is characterized by rich  saturated organic molecules like HCOOCH$_3$ and CH$_3$CH$_2$CN, which are evaporated from grain mantles.  The representative sources are IRAS 16293-2422 and NGC1333 IRAS4A/B \citep[e.g.][]{caz03, bot04, sak06}.  In these sources, carbon-chain molecules are generally deficient.  In contrast, the saturated organic molecules have not been detected so far in the WCCC sources in spite of very sensitive single-dish observations \citep[e.g.][]{sak07}.  Such a chemical variation originates from the difference in the evaporated species, and thereby in chemical composition of grain mantles.  In the WCCC mechanism, the abundances of carbon-chain molecules critically depends on the amount of CH$_4$ evaporated from grain mantles.  If the gas-phase CH$_4$ abundance is higher than the abundance of OH (typically $10^{-7}$; e.g. Herbst and Leung 1989), CH$_4$ can be a major destructor of C$^+$ instead of OH in a lukewarm region where H$_2$O is still frozen out.  Then, the WCCC occurs when the CH$_4$ abundance exceeds this level.  The CH$_4$ abundance derived for L1527 certainly fulfills this requirement, which further strengthens the WCCC picture in L1527.

In order to verify a role of CH$_4$ in the observed chemical diversity, it is important to measure the CH$_4$ abundances in various low-mass star-forming regions.  As demonstrated in this study, observations of CH$_3$D can be a useful method for this purpose.\\

\acknowledgments

We are grateful to the staff of HHT for excellent supports.  We thank Yuri Aikawa for informing us of the result of her model calculation prior to publication, and Cecilia Ceccarelli for her valuable discussions.  This study is supported by Grant-in-Aids from Ministry of Education,  Culture, Sports, Science, and Technologies of Japan (21224002 and 21740132).  Y. S. is partially supported by NSF Grant AST-1008577.




\clearpage

\clearpage

\clearpage

\begin{table}
\begin{center}
\caption{Parameters for the Observation of CH$_3$D ($J_K = 1_0 - 0_0$).\label{tab:para}}
\begin{tabular}{lc}
\tableline\tableline
Parameter&Value\\
\tableline
Telescope&HHT\\
$\eta_{\rm MB}$& 0.75\\
HPBW&32 [arcsec]\\
Rest Frequency$^{a}$& 232644.301 $\pm$ 0.075 [MHz]\\
$T_{\rm MB}^{b}$& 12 $\pm$ 2 [mK]\\
$\int T_{\rm MB} dv$($3\sigma$) & 7.4  $\pm$ 2.8 [mK km s$^{-1}$]\\
$V_{\rm LSR}^{b, c}$& 6.0 [km s$^{-1}$]\\
rms&2.2  [mK]\\
\tableline
\end{tabular}
\tablenotetext{a}{The rest frequency is taken from CDMS \citep{mul05}.  The uncertainty represents one standard deviation.}
\tablenotetext{b}{Obtained by a Gaussian fit.  In this fit, $dv$ is calculated to be (0.5 $\pm$ 0.1)~km~s$^{-1}$, but it is affected by the backend resolution (0.32~km~s$^{-1}$).  The uncertainty represents one standard deviation of the fit.}
\tablenotetext{c}{The rest frequency error ($\sim 0.1$ km s$^{-1}$) and the channel spacing of the backend (0.32 km s$^{-1}$) are larger than the Gaussian fitting error.}
\end{center}
\end{table}

\clearpage

\begin{table}
\begin{center}
\caption{Column Densities and Fractional Abundances of CH$_3$D and CH$_4$.\label{tab:column}}
\begin{tabular}{ccccc}
\tableline\tableline
Temperature&$N$(CH$_3$D) [cm$^{-2}$]$^{a}$&$f$(CH$_3$D)$^{a,b}$&$N$(CH$_4$) [cm$^{-2}$]$^{c}$&$f$(CH$_4$)$^{b, c}$	\\
\tableline
15~K	&(6.3$\pm 2.4$)$\times 10^{15}$&(2.1$\pm 0.8$)$\times 10^{-7}$&(0.9--3.1)$\times 10^{17}$&(3.0--10.5)$\times 10^{-6}$\\
25~K	&(9.1$\pm 3.4$)$\times 10^{15}$&(3.0$\pm 1.1$)$\times 10^{-7}$&(1.3--4.6)$\times 10^{17}$&(4.3--15.2)$\times 10^{-6}$\\
35~K	&(12.9$\pm 4.9$)$\times 10^{15}$&(4.3$\pm 1.6$)$\times 10^{-7}$&(1.8--6.4)$\times 10^{17}$&(6.1--21.4)$\times 10^{-6}$\\
\tableline
\end{tabular}
\tablenotetext{a}{The error is evaluated from three times the standard deviation of the integrated intensities.}
\tablenotetext{b}{The values are calculated by assuming the H$_2$ column density of $3 \times 10^{22}$~cm$^{-2}$.}
\tablenotetext{c}{The values for CH$_4$ are calculated by assuming the CH$_3$D/CH$_4$ ratio of  2--7~\% (See text).}
\tablecomments{The source size is assumed to be 20$^{\prime\prime}$ \citep{sak10}.}
\end{center}
\end{table}

\end{document}